\documentclass[12pt]{article}
\usepackage{mathpazo}
\newcommand{\ba}{\begin{eqnarray}} \newcommand{\ea}{\end{eqnarray}}
\newcommand{\nn}{\nonumber} \renewcommand{\bf}{\textbf}
\newcommand{\ra}{\rightarrow}  \newcommand{\p}{\partial}
 \newcommand{\NN}{ \nabla}

\def\bra#1{\left< #1\right|}
\def\ket#1{\left| #1\right>}

\def\EV#1#2#3{\bra{#1}#2\ket{#3}}

\begin{document}

\date{}

\title{Exploring Confinement with Spin}

\maketitle

\begin{center}

\author{ \large JOHN P. RALSTON$^{*}$  \\
{\it Department of Physics \& Astronomy \\ University of Kansas, 
Lawrence, KS 66045 USA \\
$^*$E-mail: ralston@ku.edu}}

\end{center}

\begin{abstract}
A confining gauge theory violates the completeness of asymptotic states held as foundation points of the $S$-matrix.  Spin-dependent experiments can yield results that appear to violate quantum mechanics.  The point is illustrated by violation of the Soffer bound in $QCD$. Experimental confirmation that the bound is violated would be a discovery of immense importance, sweeping away fundamental assumptions of strong interaction physics held for the past 50 years. \end{abstract}



\section{A Completeness Paradox of Confinement} 

{\it Confinement} was not anticipated in early days formulating the $S$-matrix.  Let $\ket{p, \, s}$ be a set of asymptotic hadron {\it in}-states.   The $S$-matrix elements $S_{p'p}$ are defined by \ba S_{p's';ps}= lim_{t \ra \infty} \, \EV{p', \,s'  }{e^{-i H_{tot}t}}{p,  \, s}. \nn \ea  When the exact Hamiltonian $H_{tot}$ maps states out of the Hilbert space of asymptotic states - meaning that the hadrons are not complete for their own interactions - then a classic $S$-matrix assumption is violated.  

$QCD$ poses a definite paradox with hadronic completeness.  Let $Q_{a}$ be a {\it global} color generator in equal-time quantization, \ba \ [ Q_{a},  \, Q_{b} \ ] = i f_{abc}Q_{c}, \nn \ea where $f_{abc}$ are the structure constants of the color group. Let $1_{h} =\sum_{h} \, \ket{h}\bra{h}$ be a complete set of all orthogonal hadrons of all momenta and spin. Insert the complete set, and impose they are color singlets, yielding \ba \ [ Q_{a},  \, Q_{b} \ ] = Q_{a} \,  \sum_{h} \, \ket{h}\bra{h} \, Q_{b} -Q_{b}   \sum_{h} \, \ket{h}\bra{h} \, Q_{a} \ra 0. \nn \ea  The ``paradox of colored completeness'' is {\it how do physical states support the algebra?}  Resolving this paradox leads to phenomena that superficially appear to violate quantum mechanics, while providing a fascinating probe of confinement. 

\section{Two Kinds of Gauge Transformation} 

We explore the paradox by looking more deeply into $QCD$.   Let $\psi_{a}(\,  x\,)$ be a matter field (\, ``type 1''\,) in the fundamental color representation.  Let $A_{a}( x)$ be a gauge potential (\, ``type 2''\,) in the adjoint representation. Fix $A_{0}=0$; there remains space-dependent local gauge transformations of type 1 and 2. Under a gauge transformation at equal times \ba \ [ Q_{a}^{(\, 1\,)}(\,  x\,), \, \psi_{b}(\,  x'\,) \ ]&=& i \tau_{bc}^{a}\psi_{c}\delta(\,  x- x'\,); \nn \\ \  [ Q_{a}^{(\, 2\,)}(\,  x\,), \, A_{b}(\,  x'\,) \ ]&=& i f_{abc}\delta A_{c}\delta(\,  x- x'\,), \nn \ea by which \ba     A (\,  x\,) \ra  U(\, \theta(\,  x\,)\,) \, A \,  U^{\dagger}(\, \theta(\,  x\,)\,) -i gU(\, \theta(\,  x\,)\,)\,  \p \, U^{\dagger}(\, \theta(\,  x\,)\,). \nn \ea  Here \ba \ [ Q_{a}^{(\, 1\,)}(\,  x\,), \,  Q_{b}^{(\, 1\,)}(\,  x'\,) \ ] &=& i f_{abc}Q_{c}^{(\, 1\,)}(\,  x\,)\delta(\,  x- x'\,); \nn \\ \ [ Q_{a}^{(\, 2\,)}(\,  x\,), \,  Q_{b}^{(\, 2\,)}(\,  x'\,)\ ] &=& i f_{abc}Q_{c}^{(\, 2\,)}(\,  x\,)\delta(\,  x- x'\,); \nn \\ \ [ Q_{a}^{(\, 1\,)}(\,  x\,), \,  Q_{b}^{(\, 2\,)}(\,  x'\,)\ ] &=& 0. \nn \ea  If the color group is $SU(N)$, the gauge transformations of type 1 and 2 generate $SU(\, N\,) \times SU(\, N\,)$. The breakdown of this group will explain the paradox.

\subsection{Gauss' Law}
 
Physical states are {\it gauge invariant} under the diagonal subgroup $SU(\, N\,)  \subset SU(\, N\,) \times SU(\, N\,)$ generated by $Q^{(\, 1\,)}(\,  x\,)+ Q^{(\, 2\,)}(\,  x\,) $. The invariant subspace $\ket{all}$ is defined by \ba (\, Q_{a}^{(\, 1\,)}(\,  x\,) + Q_{a}^{(\, 2\,)}(\,  x\,) \,) \ket{all} =0; \nn \\ (\,  (\, D\cdot E\,)_{a}(\,  x\,) -\rho_{a}(\,  x\,)\,) \ket{all} =0, \nn \ea where $\rho_{a}(\,  x\,)$ is the charge operator on the matter fields, and $(\, D\cdot E\,)_{a}(\,  x\,)$ is the Gauss-law operator on the gauge fields, with $D$ the gauge-covariant derivative.  Under Gauss' Law certain gluonic configurations are attached to the matter fields to make {\it singlet operators invariant under the joint transformations of type 1 and 2:} \ba \psi_{a}(\,  x\,) \otimes A_{b} \ra \psi_{a}(\,  x\,) \bar e_{a}^{\mu}(\,  x; \, A\,); \nn \\
 \ [  Q_{b}(\,  x\,), \, \psi_{a}(\,  x'\,) \bar e_{a}^{\mu}(\,  x'; \, A\,) \ ] = 0. \nn \ea Index $\mu$ (\, the ``solution number''\,) is a composite index including Lorentz properties and internal parameters. 
 
 \subsubsection{Dirac's Solution for $QED$}
 {\it For QED pure gauge-fields}, Dirac\cite{dirac} solved $e_{a}^{\mu}(\,  x; \, A\,) \ra e_{Dirac}(\,  x;\, A\,)$ \ba e_{Dirac}(\,  x;\, A\,)&= & e^{ig {\NN\cdot  A \over \NN^{2}}(\,  x\,)}. \nn \ea Under a type-2 gauge transformation $ A(\,  x\,)\ra    A(\,  x\,)+ \NN \theta(\,  x\,)$ we find: \ba  e_{Dirac}(\,  x;\, A+ \NN \theta(\,  x\,)\,) &= & e^{ig {\NN\cdot (\,  A + \NN \theta\,) \over \NN^{2}}(\,  x\,)} = e^{ig \theta(\,  x\,)} \, e_{Dirac}(\,  x;\, A\,). \nn \ea Dirac's ``dressing operator'' transforms like an electron. With $E= i\delta/\delta A$ the electric field operator, we check that Gauss' Law was solved by
\ba \EV{0}{\bar e_{Dirac}(\,  x;\, A\,) i {\delta \over \delta  A(\, y\,)} e_{Dirac}(\,  x;\, A\,)}{0} =-g {\NN \over \NN^{2}_{ x y}} . \nn \ea But Dirac's pure-gauge construction is only part of the story.  
 
\subsection{Embedded Frames and Connections}

The geometrical interpretation of $e_{a}^{\mu}$ is a {\it frame} for color polarizations on a ``big'' (embedding) space labeled with index $\mu$. By Gauss the frame $e_{a}^{\mu}$  must transform like a quark - the fundamental rep - on index $a$: \ba e_{a}^{\mu}( x) \ra U_{ab}( x)e_{a}^{\mu}( x); \:\:\:\:\: U_{ab}( x)U_{bc}^{\dagger}( x)=\delta_{ac}.  \ea
Linear combinations of solutions are expressed with index $\mu$.  The frame is ``normalized'' by its inverse $\bar e_{a \mu}$, \ba \bar e_{a \mu} e_{b}^{\mu}= \delta_{ab}. \ea The range $\mu =1...M$ depends on the geometry of the embedding space. Choosing $e_{a}^{\mu}$ a square, unitary matrix will only generate pure gauge configurations. By a theorem of John Nash one can always choose $M$ large enough so that an arbitrary manifold can be embedded in a geometrically-trivial {\it Euclidean} space. 

The {\it connection} $A_{ab}$ is \ba A_{ab}(e; \, \bar e; \,  x) = {-i \over g} \sum_{\mu} \, e_{a \mu}( x) \, \p \bar e_{b}^{\mu}( x). \nn \ea Here $\p$ means $ \p /\p x_{\beta}$. The symbol $A_{ab}$ transforms locally like type-2: \ba A_{ab}(e; \, \bar e; \, x)\ra A_{ab}(Ue; \, \bar e U^{\dagger};\, x)  =(U(x)  A(e; \, \bar e; \, x) U^{\dagger})_{ab} -{i \over g} U_{ac}(x)\, \p \,U^{\dagger}_{cb}(x). \nn \ea   The whole theory can be expressed using the frames. 
 
\subsubsection{Gauge Links and Dressed Partons}

{\it Parallel transport} of frames is described by choosing a path and an associated gauge transformation  \ba e_{P}( A; \, x, \, x') = {\cal P} e^{i g \int_{x}^{x'} \, dx \cdot A(\, z\,)}. \nn \ea Parallel transport is neither an efficient nor necessary way to set up frames.  Instead, path dependence under parallel transport is a symptom that the frames describe a curved space.  The Gauss' Law frames $e_{a}^{\mu}(x)$ and connections $A_{ab}(e; \, \bar e; \, x)$ are ordinary field that exist ``everywhere'' with no inherent path dependence.  
 
The gauge invariant coordinate-space parton distribution is  \ba \Phi(x^{-}, \, x_{T}) = \sum_{\mu} \,\EV{p, \, s}{\bar \psi_{a}(x )  e_{a}^{\mu}(x)  \Gamma \bar e_{b}^{\mu}(0) \psi_{b}(0)}{p, \, s}, \nn \ea where $\Gamma$ represents spin projections.  By Gauss' Law the frame-factors dress the bare quark operators with gluons.  Early``leading twist'' $QCD$ set $x^{+}= x_{T}=0$, implementing the impulse approximation and integrating over parton transverse momenta. The assumption of short distance trivializes dressing to an unphysical light-like line along $ x_{T}=0$. Beyond leading twist, and in discussion of initial state interactions such as the Sivers effect, the correlations of dressing with spin and momentum are just what we seek to learn from experiments. 

\section{Two Kinds of Completeness} 

We now return to the paradox of colored completeness. The full $QCD$ interaction comes into play between overlapping hadrons at finite range.  The $QCD$ Hamiltonian is only invariant under $SU(3) \subset SU^{(1)}(3) \times SU^{(2)}(3) $.  The breakdown of invariance of the type-1 and type-2 groups allows
exchange of local color and dressing while maintaining the gauge symmetry.

Consider the local generator relations \ba [ Q_{a}^{(\, 1\,)}(\, x\,), \,  Q_{b}^{(\, 1\,)}(\, x'\,) \ ] &=& i f_{abc}Q_{c}^{(\, 1\,)}(\, x\,)\delta(\, x-x'\,); \nn \\ \ [ Q_{a}^{(\, 2\,)}(\, x\,), \,  Q_{b}^{(\, 2\,)}(\, x'\,)\ ] &=& i f_{abc}Q_{c}^{(\, 2\,)}(\, x\,)\delta(\, x-x'\,) \nn \ea Let $\ket{all}$ be a complete set of {\it all} states of hadrons under interactions. Inset $1_{all} = \sum_{all} \, \ket{all}\bra{all}$, giving \ba  [(\, Q_{a}^{(\, 1\,)}(\, x\,) &+& Q_{a}^{(\, 2\,)}(\, x\,)  \,) \, \sum_{all} \, \ket{all} \bra{all} \, (\,  Q_{b}^{(\, 1\,)}(\, x'\,)+ Q_{b}^{(\, 2\,)}(\, x' \,) \,) \ ] \nn \\  &=& i f_{abc}Q_{c}^{(\, tot\,)}(\, x\,)\delta(\, x-x'\,); \nn \\ Q^{(tot)}(x) & =& Q_{a}^{(\, 1\,)}(x)+ Q_{a}^{(\, 2\,)}(x).  \nn \ea We made it explicit that interacting states transform under the type-1 and type-2 groups. But these transformations are not good symmetries, so we do not get new quantum numbers. In comparison the asymptotic $\ket{p, \, s}$ only transform under the Lorentz group and global symmetries that commute.  It follows that the two spaces $\ket{p, \, s}$ and $\ket{all} $ are not equivalent.  Completeness of $\ket{all}$ is ``bigger'' than completeness of trivial asymptotic states (lacking color labels):  \ba 1_{h} =\sum_{h} \,  \ket{h}\bra{h} \subset 1_{all} = \sum_{all} \,  \ket{all}\bra{all}    . \nn \ea  This resolves the paradox.

\section{Reduction} 
Without much discussion parton phenomenology implements the more-complete completeness of interacting states by giving color indices to perturbative quark and gluon lines at intermediate steps in the calculation. Summation over the color indices produces {\it reduction}, a step decreasing the dimension of a direct-product space. 

Reduction requires density matrix formalism. It is convenient to suppress all dimensions except color and spin. Let $\ket{\chi_{\alpha}}$ be a basis element on the spin space, and $ \ket{\phi_{i}}$ be a basis element on the color space, which spans type-1 and type-2.  Take a particular state $\ket{\psi }$ on the interacting space, and 
expand: \ba \ket{\psi} = \sum_{i \alpha} \ket{\chi_{\alpha}} \ket{\phi_{i}} \, \psi_{i \alpha}, \label{expand} \ea Any pure state $\ket{\psi}$ is equivalent to a density matrix $\rho_{\psi}$ made from the outer product with itself.  The density matrix on asymptotic states comes from tracing out the color: \ba \rho_{h}&  =&  tr_{color} (\rho) = tr_{color} (\ket{\psi}\bra{\psi}) =\sum_{i}  \, \ket{\chi_{\alpha}} \psi_{i\alpha}\psi_{i \beta}^{*} \bra{\chi_{\beta}}. \nn \ea After reduction it is not generally possible to represent $\rho_{h}$ as the outer product of a pure state with itself.  Even more interesting, $\rho_{h}$ will not describe hadron dynamics when color is exchanged. {\it Hadrons are incomplete to describe their own interactions}

Transitions of a general operator $\Omega$ are given by \ba S = tr(\rho_{all, \, in}\Omega\rho_{all, \,out}\Omega). \nn \ea  Since neither $\rho_{h}$ nor $\ket{p, \, s}$ appear, properties of superposition assuming a wave function can fail.   It is not quantum mechanics that is violated, however, it is the naive use of quantum mechanics developed for pure states that is violated. 

I illustrate this by showing that the Soffer bound can be violated in $QCD$.  

\section{The Soffer Bound} 

Using the rules of quantum mechanics for pure states Soffer\cite{soffer} produced a bound on $h_{T}(x, \, Q^{2}) $, the distribution of transversely polarized quarks in a transversely polarized target\cite{john}.  Soffer's bound is \ba h_{T}(x, \, Q^{2}) < |q(x, \, Q^{2})+ \Delta q(x, \, Q^{2})|/2, \nn \ea where $q(x, \, Q^{2})$ and $\Delta q(x, \, Q^{2}$ are the unpolarized and longitudinally polarized quark distributions.  Soffer's bound is more restrictive than positivity. A recent review by Artru {\it et al}\cite{artru} presents the bound and many related spin inequalities.  Barone {\it et al} \cite{Barone:2001sp} handsomely review the general subject of transverse polarization. Experiments are in flux, but indicate that transversity is large\cite{Anselmino:2007fs}. The relation of experiments to matrix elements is the important goal of the transversity project reviewed in these Proceedings.  Let us re-examine the use of quantum mechanics in the reduced system. 

Eq. \ref{expand} can be simplified using the singular value decomposition: \ba \ket{\psi} = \sum_{\alpha}\, \ket{\tilde \chi_{\alpha}} \Lambda_{\alpha} \ket{\tilde \phi_{\alpha}} . \label{svd} \ea The tilde-basis are orthonormal, and can be found from diagonalizing two reduced density matrices on color and spin. Notice that the number of terms cannot exceed the dimension of the smaller space.  The interpretation of Eq .\ref{svd} is a basis in which each tilde-spin and color state  are strictly correlated, with weights represented by $\Lambda_{\alpha}$  Correlation of color and spin would ordinarily seem to violate gauge invariance, but we have done the work to show it is an outcome of gauge invariance.  The tilde-basis is particular to the particular state $\ket{\psi}$, which is time-dependent, and dynamical.

Consider an asymptotic spin 1/2 hadron helicity basis $\ket{h\pm}$, for which transversely polarized states $\ket{h T_{\pm}}$ are related by ordinary superposition: \ba \ket{h T_{\pm}} = {1 \over \sqrt{2}}(\ket{h+} \pm \ket{h-}). \nn \ea A measurement in a ``pure '' transversely polarized state is described by \ba P(\Omega) &=& tr( \ket{T_{+}}\bra{T_{+}} \Omega), \nn \\ &=& {1\over 2}( tr( \ket{ h+}\bra{h+} \Omega)+ tr( \ket{ h+}\bra{h-} \Omega)+ ...\nn \ea Soffer's bound comes by letting $\Omega$ be the correlations of transversely polarized quarks. 
In reality an experiment measures \ba P(\Omega)=\sum_{\alpha, \beta} \, \Lambda_{\alpha} \Lambda_{\beta}(  \bra{\tilde \phi_{\alpha}}\bra{\tilde \chi_{\alpha}} )\,  \Omega \,  ( \ket{\tilde \chi_{\beta}}   \ket{\tilde \phi_{\beta}}). \nn \ea It is not possible to proceed without detailed information about the time evolution in $QCD$. 

Some relations can by developed by making unrealistic restrictions.  Suppose an operator does not correlate color with spin: $\Omega \ra \Omega_{color} \otimes \Omega_{spin}$. Then \ba <\Omega>& \ra & tr(\rho_{h; \, \Omega} \Omega_{spin}); \nn \\ \rho_{h; \, \Omega}& =& \sum_{\alpha} \, \ket{\tilde \chi_{\alpha}(t)} \rho_{\alpha \beta }(t) \bra{\tilde \chi_{\alpha}(t) }, \nn \\ & where & \:\:\:\: \rho_{\alpha \beta}(t) \ = \Lambda_{\alpha}(t) \EV{\tilde \phi_{\alpha}(t)}{\Omega_{color}}{ \tilde \phi_{\beta }(t)} \Lambda_{\beta}(t) .  \nn \ea  In this case an effective hadronic theory exists, but it is driven from outside by the color interactions the same theory cannot express.  Requiring that the interacting hadronic system remain pure at all times would be completely arbitrary. $QCD$ has no such rules, so we again conclude that hadrons are incomplete for their own interactions. For instance chiral models, which are expressed entirely by hadronic degrees of freedom, can't possibly represent $QCD$, which happens to be consistent with the fact nobody uses chiral models at high energies.  It is much more surprising to realize {\it there is not supposed to be a local effective hadronic theory of any kind} representing $QCD$.

Soffer's bound fails because the final state ``complete sets'' depend on how the system was prepared and interacts.  We can be quite sure that the final state hadrons of a transversely polarized reaction will differ from those of  longitudinally polarized one. One would expect detailed calculations to reveal the general effect.  The $Q^{2}$ dependence of parton distributions calculated perturbatively confirms the general argument. The anomalous dimensions of $h_{T}, \, \Delta q$, and $q$ have been calculated to leading and next to leading order\cite{Efremov:1983eb}. Since the functions scale by different rules, Soffer's bound must fail under evolution in one direction (increasing $Q^{2})$ or the other (decreasing $Q^{2})$.  Neither outcome is acceptable under the strict logic used to make the bound. Consistency problems were noticed before\cite{violated}.  However the bound was compared to the Callan-Gross relation, which is a correct zeroth order kinematic relation disturbed by radiative corrections.  Our statement here is much stronger: 
the Soffer bound is based on premises that became obsolete with confintement.  Examining the size and the direction of $pQCD$ effects is a side issue that does not address faults of the non-perturbative starting point. 
 
The breakdown of hadronic completeness is the central issue.  It would be good if studies extracting transversity distributions would assume {\it positivity}, rather than the Soffer bound, in order not to prejudice the analysis,  Experimental confirmation that the bound is violated would be a discovery of immense importance, sweeping away fundamental assumptions of strong interaction physics held for the past 50 years.

\section{Acknowledgments } 
I thank the workshop organizers for their generosity, and Mauro Anselmino, Leonard Gamberg, Gary Goldstein,  Kloya Nikolaev and Oleg Teryaev for helpful discussions.  Research supported in part under DOE Grant Number DE-FG02-04ER14308.

\end{document}